\documentclass[12pt]{iopart}
\usepackage{graphicx}
\usepackage{iopams}

\newcommand{\prb}{{\em Phys. Rev. B~}}
\newcommand{\physc}{{\em Physica C~}}

\newcommand{\xhat}{\hat{\textbf{x}}}
\newcommand{\yhat}{\hat{\textbf{y}}}
\newcommand{\zhat}{\hat{\textbf{z}}}
\newcommand{\bk}{\textbf{k}}
\newcommand{\ka}{\textbf{k}_1}
\newcommand{\kb}{\textbf{k}_2}

\newcommand{\q}{\textbf{q}}

\newcommand{\cis}{c_{i,\sigma}}
\newcommand{\cjs}{c_{j,\sigma}}

\newcommand{\cisp}{c^{\dag}_{i,\sigma}}

\newcommand{\cmup}{c^{\dag}_{m,\uparrow}}

\newcommand{\cndp}{c^{\dag}_{n,\downarrow}}

\newcommand{\vac}{|0\rangle}

\begin{document} 

\title[Bound state formation in exchange-coupled planes]
{Two-electron bound state formation in the $t-J-U$ model for exchange-coupled planes}

\author{A. Morriss-Andrews and R. J. Gooding}
\address{Department of Physics, Queen's University, Kingston ON K7L 3N6
Canada}


\begin{abstract}

An anisotropic $t-J-U$ model Hamiltonian is used to model electron behaviour in quasi-2$d$
materials in the dilute limit, and as
a highly simplified representation of the weakly coupled CuO$_2$ planes of the high-$T_c$ cuprates
we model the very poor out-of-plane conductivity via the complete suppression of interplanar
hopping. However, we do include the very weak interplanar superexchange, and are thus considering a
model of exchange-coupled planes. For an isotropic three-dimensional system in the dilute limit, 
we find that the formation of two-particle bound states requires $J_c/t \gtrsim 5.9$. Also,
it is known that $J_c/t = 2$ for a 2$d$ square lattice. However, for our model of exchange-coupled 
planes any infinitesimal interplanar exchange ($J_c = 0$) is adequate to form bound states.\\ \\
Published as J. Phys.: Condens. Matter {\bf 19}, 386216 (2007).

\end{abstract}


\maketitle

\section{Motivation and Introduction} \label{intro}

The discovery of high-temperature superconductivity in copper-based transition metal oxides
was made by Alex M\"{u}ller and Georg Bednorz in 1986 (see \cite{HTCwebsite} for more information). 
They found that the copper oxide compound $(La,Ba)_2CuO_4$ became a superconductor at 
$T_c \approx 30 K$, which was substantially higher than for other any other compounds known 
at that time. Presently, one can find transition temperatures at ambient pressure close to 
138K for other cuprate-based systems.
The mechanism for this novel behaviour is still a subject of spirited debate, but one
idea that has been put forward repeatedly is the Heisenberg superexchange between
neighbouring Cu sites, a biproduct of the strong repulsive electronic (Hubbard-type)
correlations that are present on the transition metal sites.

The structure of all of the cuprate materials is similar, in that $CuO_2$ planes are stacked one
on top of another to produce a quasi-2$d$ crystallographic arrangement. This characterization
of these materials is supported by their very poor interplanar conducting 
behaviour~\cite{giura}, at least in the weakly doped regime~\cite{lai,zha}. The so-called\break 
$c$-axis puzzle has been studied extensively \cite{ginsberg}, and its relation to
the superconductivity has been discussed \cite{pwa}.
Further, given the dependence of $T_c$ on the number of $CuO_2$ planes per unit 
cell~\cite{kuzemskaya}, the possibility of
an interplanar pairing mechanism cannot be ignored, and indeed previous work has 
shown~\cite{odonovan,mraz}, within various approximations, that any interplanar interaction 
increases $T_c$.

In this report we examine the two-electron problem (\textit{viz.} the dilute limit) in the 
highly simplified situation of zero interplanar hopping. That is, our model is meant to be 
a very rough approximation to the extremely low out-of-plane conductivity, but leaving the 
residual interplanar superexchange found in the cuprates, the latter produced by 
strong electronic correlations. We refer to this situation as \textit{exchange-coupled planes}.
We solve for 
the conditions under which a two-particle bound state is formed, and in particular determine
the minimum value of superexchange ($J_c/t$) for which bound states appear. 
Our results make evident
the potential benefit of having electrons confined to individual planes that can only
move via the spin-exchange ($\frac{J}{2}S^+_iS^-_j$) process.  To be specific, we find
that while for a two-dimensional plane ($J_c/t = 2$) or an isotropic three-dimensional
system ($J_c/t \approx 5.9$) one requires a superexchange much larger than that found
experimentally ($J/t \sim 0.3$), for exchange-coupled planes one requires only an
infinitesimal ($J_c =0$) interaction. Therefore, even the very small out-of-plane
exchange coupling (estimated to be roughly $10^{-4}$ of the in-plane exchange) would
suffice to form bound states. We note that similar
results were found previously for exchange-coupled chains~\cite{basu}, emphasizing the
potential importance of such \textit{electronic confinement}. 

\section{Formalism}
\subsection{Model Hamiltonian}

We consider the so-called $t-J-U$ or Heisenberg-Hubbard model Hamiltonian, given by
\begin{equation} 
\label{eqnH}
{\cal H} = - \sum_{\left\langle i,j\right\rangle,\sigma}(t_{i,j}\cisp\cjs + h.c) 
+ \sum_{\left\langle i,j\right\rangle}J_{i,j} (\textbf{S}^{(i)} \cdot \textbf{S}^{(j)} 
- \frac{1}{4}n_{i} n_{j}) + U \sum_i n_{i\uparrow}n_{i\downarrow}\quad
\end{equation}
In this Hamiltonian $i,j$ label the sites of the lattice, the notation $\langle i,j\rangle$ 
denotes neighbouring lattices sites in \textit{any} of the $x,~y$ or $z$ directions, each near
neighbour pair is counted once only, and $\cis/\cisp/n_{i,\sigma}$ denotes the 
annihilation/creation/number operator for an electron at site $i$ with spin $\sigma$.
We will consider the dilute limit, and in particular only two electrons, and therefore
the frustrating geometry of some the high-$T_c$ cuprates (such as
occurs for a body-centred tetragonal structure) can be ignored. Therefore, 
we consider a simple tetragonal lattice structure, which reduces to a simple cubic
structure for isotropic hopping and exchange -- see below.

\begin{figure}
\begin{center}
\includegraphics[height=2.5in]{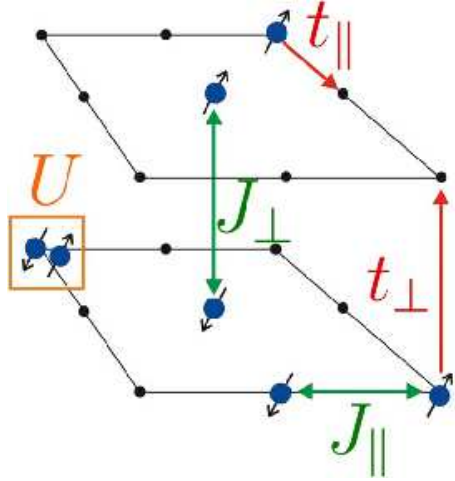}
\end{center}
\caption{(Colour online) A schematic representation of the various terms that appear in the $t-J-U$ model
Hamiltonian. The electron hopping $t$, both in-plane ($t_\parallel$) and inter-plane ($t_\perp$), 
spin exchange $J$, again for both in-plane ($J_\parallel$) and inter-plane ($J_\perp$), and 
the on-site Coulomb repulsion $U$.}
\label{fig:nrgparams}
\end{figure}

For the energy parameters in the Hamiltonian we consider the following situations; a schematic
of these parameters is shown in figure~\ref{fig:nrgparams}. First,
the hopping frequency is restricted to be near neighbour only, and in-plane and
out-of-plane hopping frequencies are allowed to be different:
\begin{eqnarray}
\label{eq:tijs}
t_{i,j}&=&t_\parallel\hspace{1.0cm}\textrm{when}~i,j~\textrm{are near neighbours in either the}
~x~\textrm{or}~y~\textrm{directions}
\nonumber \\
&=&t_\perp\hspace{1.0cm}\textrm{when}~i,j~\textrm{are near neighbours in the}~z~\textrm{direction}
\end{eqnarray}
Similarly for the superexchange integral:
\begin{eqnarray}
\label{eq:Jijs}
J_{i,j}&=&J_\parallel\hspace{1.0cm}\textrm{when}~i,j~\textrm{are near neighbours in either the} 
~x~ \textrm{or}~\textrm{y directions}
\nonumber \\
&=&J_\perp\hspace{1.0cm}\textrm{when}~i,j~\textrm{are near neighbours in the}~z~\textrm{direction}
\end{eqnarray}
The Hubbard on-site repulsion energy is parametrized by $U$. 

Strong coupling expansions of the Hubbard model show that to lowest order in $t/U$ one has $J=4t^2/U$,
and in the $t-J-U$ model this exchange interaction is included explicitly.
Note, however, that this model Hamiltonian does \textit{not} involve projection operators that project 
away doubly occupied sites, as one considers in $t-J$ model studies~\cite{dagotto94}. 
Instead, we take the limit $U\rightarrow\infty$ to represent the high cost of doubly occupied sites,
similar to what has been done in previous studies~\cite{basu,lin}.
That is, $J$ is the effective interaction that is created by some finite value of $U$~\cite{gros},
and we then reduce the Hilbert space by taking the $U\rightarrow \infty$ limit leaving $J$ nonzero. 

As we have discussed in the introduction, in this paper we compare the bound states for a two-electron 
system, described by the Hamiltonian of (\ref{eqnH}), when interplanar hopping does not 
occur (that is $t_z=0$), to the situation in each of the hopping and superexchange energies 
in (\ref{eq:tijs}) and (\ref{eq:Jijs}) are the same in all three directions.

Consider the two-electron, $S^z_{\textrm{total}} = 0$ state
\begin{equation} \label{eqnPsi}
        |\psi\rangle = \sum_{m,n} \phi(m,n) \cmup \cndp \vac
\end{equation}
created from the empty lattice, $\vac$. With the added symmetry $\phi(m,n) = \phi(n,m)$, $|\psi\rangle$ will be a
singlet state.
We use the following for the Fourier coefficients for $\phi(i,j)$: 
\begin{equation}
\phi(\ka,\kb) \equiv \frac{1}{N} \sum_{i,j} e^{-i(\ka\cdot\textbf{r}_i+\kb\cdot\textbf{r}_j)}\phi(i,j)
\end{equation}
Introducing $\ka=\frac{\bf Q}{2}+{\bf q}$ and $\kb=\frac{\bf Q}{2}-{\bf q}$, we restrict our attention
to zero centre-of-mass momentum states, and therefore solve for solutions in terms of
\begin{equation}
\phi(\ka,\kb) = \phi({\bf q},-{\bf q})\equiv\phi({\bf q})
\end{equation}

We make the following definitions:
\begin{eqnarray}  
\label{eqnt}
\epsilon(\bk)&\equiv& -2 \{t_\parallel [\cos (\bk\cdot \xhat) + \cos (\bk\cdot \yhat)] + t_\perp \cos (\bk\cdot \zhat)\}
\\
\label{eqnJ}
J(\bk)&\equiv& 2 \{J_\parallel[\cos (\bk\cdot \xhat) + \cos (\bk\cdot \yhat)] + J_\perp \cos (\bk\cdot \zhat)\}
\end{eqnarray}  
and find~\cite{lin,basu} that $H|\psi\rangle=E|\psi\rangle$ is equivalent to solving the following equation:
\begin{equation} \label{eqnMain}
\phi(\q) = \frac{\frac{U}{N} \sum_\textbf{p} \phi(\textbf{p}) - \frac{1}{N}\sum_\textbf{p} J(\q - \textbf{p})\phi(\textbf{p})}{E - 2\epsilon(\q)}
\end{equation}

To simplify (\ref{eqnMain}) it is convenient to make the following definitions 
(with $x_i$ and $x_j$ being any of $x, y$ or $z$):
\begin{eqnarray} \label{eqnC}
	C_0 &\equiv& \frac{1}{N} \sum_\bk \phi(\bk) \quad\quad\quad
	C_{x_i} \equiv \frac{1}{N} \sum_\bk \cos(k_{x_i}) \phi(\bk) \\
	\nonumber \\
	\label{eqnI}
	I_0 &\equiv& \frac{1}{N}\sum_\q \frac{1}{E + 4(t_\parallel[\cos(\q \cdot \xhat) + \cos(\q \cdot \yhat)] + t_\perp \cos(\q \cdot \zhat))} \\
	I_{x_i} &\equiv& \frac{1}{N}\sum_\q \frac{\cos(\q \cdot \hat{\textbf{x}}_i)}
{E + 4(t_\parallel[\cos(\q \cdot \xhat) + \cos(\q \cdot \yhat)] + t_\perp \cos(\q \cdot \zhat))} \nonumber \\
	I_{x_i,x_j} &\equiv& \frac{1}{N}\sum_\q \frac{\cos(\q \cdot \hat{\textbf{x}}_i) \cos(\q \cdot \hat{\textbf{x}}_j)}{E + 4(t_\parallel[\cos(\q \cdot \xhat) + \cos(\q \cdot \yhat)] + t_\perp \cos(\q \cdot \zhat))} \nonumber
\end{eqnarray}
By substituting these definitions into (\ref{eqnMain}) one straightforwardly obtains a system of equations 
that one can solve for the bound states of the problem:
\begin{eqnarray} \label{eqnSet}
	C_0 &=& U I_0 C_0 - 2 J_\parallel(I_x C_x + I_y C_y) - 2 J_\perp I_z C_z \\
	C_x &=& U I_x C_0 - 2 J_\parallel(I_{xx} C_x + I_{xy} C_y) - 2 J_\perp I_{xz} C_z \nonumber \\
	C_y &=& U I_y C_0 - 2 J_\parallel(I_{xy} C_x + I_{yy} C_y) - 2 J_\perp I_{yz} C_z \nonumber \\
	C_z &=& U I_z C_0 - 2 J_\parallel(I_{xz} C_x + I_{yz} C_y) - 2 J_\perp I_{zz} C_z \nonumber 
\end{eqnarray}

\section{Results}

We analyzed the above system of equations for two geometries of lattices: infinite in all directions 
(referred to as the \textit{3$d$ case}), as well as for $L_x = L_y = \infty$ and $L_z=2$ (referred to 
as the \textit{two plane problem}). For the 3$d$ case, where $L_x,L_y,L_z \rightarrow \infty$, the 
sums in (\ref{eqnC}) and (\ref{eqnI}) become
\begin{equation} \label{eqnIInt}
\frac{1}{N}\sum_\q \rightarrow \frac{1}{(2\pi)^3}\int^\pi_{-\pi} dq_x\int^\pi_{-\pi} dq_y\int^\pi_{-\pi} dq_z 
\end{equation}
and similarly for an isotropic 2$d$ lattice.
Also, for $L_x,L_y \rightarrow \infty$ in the two plane problem ($L_z=2$) the sums become
\begin{equation}
\frac{1}{N}\sum_\q \rightarrow \frac{1}{2} \sum_{m_z = 0,1}\frac{1}{(2\pi)^2}\int^\pi_{-\pi} dq_x\int^\pi_{-\pi} dq_y
\end{equation}

For the $t-J-U$ model with isotropic hopping ($t$) and exchange ($J$) in dimension $d$, 
the minimum (total) energy 
for two noninteracting electrons is -4$dt$, and therefore the energy of a bound state must be less than 
this energy. For convenience, we define a scaled energy $A$ by
\begin{equation}
	A \equiv \frac{E}{4 t}
\end{equation}

One of our results will be related to the simpler and lower-dimensional isotropic single-plane problem, 
and thus here we outline a method of solving such a problem. Eliminating the $z$
dependence and using the equivalence of the $x$ and $y$ directions, (\ref{eqnSet}) reduces to
\begin{eqnarray} \label{eqnSet2DSIMPLE}
	C_0 &=& U I_{0,2d} C_0 - 4 J I_{x,2d} C_x \\
	C_x &=& U I_{x,2d} C_0 - 2 J I_{xx,2d} C_x - 2 J I_{xy,2d} C_x \nonumber 
\end{eqnarray}
One can further simply these equations using the identities~\cite{lin}
\begin{eqnarray} \label{eqnI2D}
	I_{x,2d} &=& \frac{1}{8t} - \frac{A I_{0,2d}}{2} \\
	I_{xx,2d} &=& - I_{xy,2d} - A I_{x,2d} \nonumber
\end{eqnarray}
and one can integrate $I_{0,2d}$ to obtain
\begin{equation} \label{eqnInt2D}
	I_{0,2d} = \frac{1}{2\pi t A} K(2/A)
\end{equation}
where $K(x)$ is the complete elliptic integral of the first kind. Thus $I_{0,2d}$ is a symmetric function 
of $A$, which is defined for $|A| > 2$ and diverges as $|A| \rightarrow 2^+$.
Now we eliminate $C_0$ and $C_x$ from the equation set (one can show that no solutions exist for either of
these being zero -- see below for 3$d$), and then take the strong Coulombic repulsion limit 
of $U \rightarrow \infty$, and then solve for $J$. One finds
\begin{equation} \label{eqnJ2d}
	J = \frac{-16 I_{0,2d} t^2}{-1+4 A I_{0,2d} t}
\end{equation}
Noting the dependence of $I_{0,2d}$ on energy, this equation (and we will derive similar equations for 
3$d$ below) then solves for the bound state energy in terms of $J$. Alternatively, one can use
this equation to solve for the minimum value of $J/t$ that is required for a bound state to form.
That is, by using (\ref{eqnInt2D}) one finds that 
as $A\rightarrow -2^-$ $I_{0,2d}$ diverges and the limiting value of $J/t$ is two. 
Hence a critical value of $J_c/t = 2$ 
is obtained in the 2$d$ isotropic case, which is in agreement with the results obtained 
elsewhere~\cite{basu,lin,pethukov}.

\subsection{Three Dimensions - Isotropic Hopping}

The elliptic integrals of (\ref{eqnI}) become impossible to solve analytically in higher dimensions. 
However, to make progress with the three-dimensional problem it is necessary to understand as much about 
the properties of these integrals as possible. $I_{0,nd}$ is an odd function that is only defined for $|A| > n$ and it is negative for $A < -n$. It is an important fact that the behaviour of $I_{0,nd}$ is qualitatively different 
for $n<3$ than for $n\geq 3$. This is because the limit as $A \rightarrow -n^-$ is divergent for $n<3$  but
finite for $n\geq 3$, a result that can be obtained, \textit{e.g.}, by writing $I_{0,nd}$ in 
$n$-dimensional polar coordinates. 

To get a simple expression of $J$ as a function of the energy $A$ in three dimensions, analogous to 
(\ref{eqnJ2d}), one can derive a set of equations analogous to (\ref{eqnI2D}), \textit{viz.}
\begin{eqnarray} \label{eqnI3d}
	I_{x,3d} &=& \frac{2}{3} \left(\frac{1}{8t} - \frac{A I_{0,3d}}{2}\right) \\
	I_{xx,3d} &=& - A I_{x,3d} - 2 I_{xy,3d} \nonumber
\end{eqnarray}

Here we analyze the infinite 3$d$ lattice with completely isotropic hopping frequencies 
and spin exchanges: $J = J_\parallel = J_\perp$ and $t = t_\parallel = t_\perp$ with $L_x = L_y = L_z = \infty$. 
Starting from (\ref{eqnSet}), and accounting for the symmetries in the elliptic integrals 
($I_x = I_y = I_z, I_{xy} = I_{yz} = I_{xz}, I_{xx}=I_{yy}=I_{zz}$ -- note that below, for simplicity,
we refrain from using the 3$d$ label for these integrals), one can eliminate 
$C_x, C_y$ and $C_z$ from the equation set to obtain the equation:
\begin{eqnarray}
\label{eq:21}
(-2I_{xx}J+2I_{xx}JUI_{0}-6UI_{x}^2J-4JI_{xy}+4JI_{xy}UI_{0}-1+UI_{0})C_0 = 0\quad\quad
\end{eqnarray}

First we assume that $C_0 \neq	0$ and solve for $U$:
\begin{equation} \label{eqnU3D}
U = \frac{2I_{xx}J+1+4JI_{xy}}{2I_{xx}JI_{0}-6I_{x}^2J+4JI_{xy}I_{0}+I_{0}}
\end{equation}
For finite $A$ and $J$, the numerator of this expression is bounded --
the $I$ integrals are convergent for $|A|>3$ and finite. 
Thus taking the limit that $U \rightarrow \infty$ is equivalent to setting the denominator to zero:
\begin{equation}
2I_{xx}JI_{0}-6I_{x}^2J+4JI_{xy}I_{0}+I_{0} = 0
\end{equation}
and then solving for $J$ gives:
\begin{equation}
J = -\frac{1}{2}\frac{I_{0}}{I_{xx}I_{0}-3I_{x}^2+2I_{xy}I_{0}}
\end{equation}
Finally, (\ref{eqnI3d}) can be used to give the simplest form for $J$, similar to (\ref{eqnJ2d}),
which again only depends on the elliptic integral $I_0$:
\begin{equation} \label{eqnJc}
J = \frac{24 I_0 t^2}{-1 + 4 A I_0 t}
\end{equation} 
The right-hand side of equation (\ref{eqnJc}) is a monotonically decreasing function as 
$A$ approaches $-3$ from below, and therefore the minimal value of $J$ is found from this limit.
One can evaluate the right-hand side of this equation numerically to a very high accuracy.
For example, one can complete the lattice sums of equation (\ref{eqnI}) with larger and
larger lattices in the limit of $E$ approaching the bottom of the noninteracting band, that is
-12$t$. However, a more accurate determination of this quantity can be obtained from
the integral form of $I_0$, which, using the properties of elliptic integrals to simplify
the multi-dimensional integral to an integration over a single variable, is given by
\begin{equation}
I_0~=~\frac{1}{2\pi}~\int_{-\pi}^{\pi} \frac{K(2/(A+\cos q))}{(2\pi)^2t(A+\cos q)}~dq
\end{equation}
where $K(x)$ is the complete elliptic integral of the first kind.
Again, examining the limit of $A$ approaching -3 from below, and carefully accounting
for the properties of the integrand near $q=0$, one finds, in
agreement with the lattice sum method mentioned above,
that $I_0 (A\rightarrow -3) = (-0.12633 \pm 0.00009)/t$. Therefore,
for this system we find $J_c/t = 5.877 \pm 0.008$. 

As mentioned, in the above derivation we have assumed for this solution that $C_0 \neq 0$ in writing 
down (\ref{eqnU3D}). Putting $C_0 = 0$ into the initial equation set, (\ref{eqnSet}) gives 
us either the trivial solution $C_0 = C_x = C_y = C_z = 0$ or a new equation for $J$. 
First we set $C_0 = 0$ in (\ref{eqnSet}), and eliminating $C_x,~C_y,~C_z$ from the equations yields
\begin{equation} \label{eqnJcAlt}
J = \frac{12 t}{-A + 4 A^2 I_0 t - 36 I_{xx} t}
\end{equation} 
Numerical studies also indicate that this $J$ has a similar qualitative behaviour as (\ref{eqnJc}), 
so if the limit as $A \rightarrow -3^-$ of (\ref{eqnJcAlt}) is less than $5.877$ then the initial 
calculation of $J_c = (5.877 \pm 0.008)t$ will be supplanted by the lower value. However, numerical 
calculations give a limit of $J$ in (\ref{eqnJcAlt}) of about 10.7.
We note that this solution has a different symmetry than the first, a result that follows
immediately from $C_0=0$; so, this latter bound state is not an on-site or extended $s$-wave solution 
(see \cite{pethukov} for a discussion of analogous results in 2$d$).

Finally, the critical spin exchange in the isotropic 3d case is determined to be 
$J_c = (5.877 \pm 0.008)t \neq 2t$. In part, we have provided this detailed analysis here because
our results contradict a statement in \cite{lin} that the 
value of $J_c/t=2$ (in the isotropic case) is independent of the lattice size and dimensionality. 

\subsection{Two planes with no $z$ hopping}\label{sub2PlaneNoHop}

We now consider the anisotropic case of no $z$ hopping in the lattice, for the reasons already reviewed
in the introduction to this paper. Physically, one can see that to study an electron pair with no $z$ 
hopping then one need only study the two plane problem since if the electrons are to see each other at 
all, they must be either on the same plane, or on adjacent $z$ planes where they may interact via the 
spin exchange interaction $J_\parallel$ or $J_\perp$. We only consider a single electron pair, and they are restricted to be in two neighbouring planes,
so the problem collapses entirely to the two plane problem. Further, due to this restricted hopping
the bottom of the two-electron scattering continuum will correspond to $E=-8t$.

We start with the two plane problem, $L_x = L_y = \infty$ and $L_z = 2$, 
that is isotropic in $x$ and $y$ and has zero $z$-hopping frequency. 
Here, as in (\ref{eqnC})--(\ref{eqnSet}) set $J_\parallel = J$. 
(The fact that for a spatial extent of $L_z=2$ it is necessary to specify open
or periodic boundary conditions does not affect our results -- see below.) Also, for the hopping
we take $t_\parallel = t$ and $t_\perp = 0$. A great deal of simplification may be made with the $I$ 
integrals of (\ref{eqnI}) because the denominator of the integrand is independent of $q_z$ 
(and $z$ is the only coordinate exhibiting anisotropy), and we can relate these integrals (which we 
denote by $\tilde I$) to their counterparts in the two-dimensional fully isotropic case. For example:
\begin{eqnarray}
	{\tilde I}_{zz} &=& \frac{1}{2L_x L_y} \sum_{m_z=0,1} \sum_{q_x,q_y} \frac{\cos^2(2\pi m_z/2)}{E+4t(\cos(q_x)+\cos(q_y))} \\
	&=& \frac{1}{L_x L_y} \sum_{q_x,q_y} \frac{1}{E+4t(\cos(q_x)+\cos(q_y))} \nonumber \\
	&=& I_{0,2d} \nonumber
\end{eqnarray}
Similarly
\begin{eqnarray} \label{eqn2DRedux}
	{\tilde I}_{0} &=& I_{0,2d} \\
	{\tilde I}_{\alpha}&=& I_{\alpha, 2d} \textrm{~~~where } \alpha = x,y  \nonumber \\
	{\tilde I}_{\alpha,\beta}&=& I_{\alpha,\beta, 2d} \textrm{~where } \alpha,\beta = x,y  \nonumber \\
	{\tilde I}_{z}&=& 0  \nonumber \\
	{\tilde I}_{\alpha z}&=& 0 \textrm{~~~~~~~where } \alpha = x,y  \nonumber \\
	{\tilde I}_{zz} &=& I_{0,2d} \nonumber
\end{eqnarray}
With these simplifications, (\ref{eqnSet}) simplifies significantly:
\begin{eqnarray} \label{eqnSet2PlaneNoHop}
	C_0 &=& U I_{0,2d} C_0 - 2 J I_{x,2d} C_x - 2 J I_{y,2d} C_y \\
	C_x &=& U I_{x,2d} C_0 - 2 J I_{xx,2d} C_x - 2 J I_{xy,2d} C_y \nonumber \\
	C_y &=& U I_{y,2d} C_0 - 2 J I_{xy,2d} C_x - 2 J I_{yy,2d} C_y\nonumber \\
	C_z &=&  - 2 J_\perp I_{0,2d} C_z \nonumber 
\end{eqnarray}

What is interesting about this set of equations is that the first three equations are identical to the 
2$d$ isotropic case, so for $C_0 \neq 0$ we find that (\ref{eqnJ2d}) holds here as well. We can 
satisfy the third equation simply by setting $C_z = 0$. Since the $z$ spin exchange is completely 
irrelevant to the equation, this physically means that this solution puts both electrons on the 
\textit{same} plane, and thus the problem reduces completely to the 2$d$ isotropic case. 

However, in our two-plane problem we can also obtain an equation for (in this case) $J_\perp$ by allowing 
$C_x = C_y = C_0 = 0$, and not get a trivial zero solution to the eigenvector of the Hamiltonian by 
having $C_z \neq 0$. Hence from (\ref{eqnSet2PlaneNoHop}) we immediately obtain:
\begin{equation} \label{eqnJz}
	J_\perp = -\frac{1}{2}\frac{1}{I_{0,2d}}
\end{equation}
In this case $J_\parallel$ has dropped completely out of the equations, and this case corresponds to 
putting the two electrons on different planes, so they can only interact through $J_\parallel$. 
Recalling that $I_{0,2d}$ diverges this gives us a critical spin coupling constant of 
$J_c = J_{\parallel,c} = 0$. That is, we may obtain bound state solutions for arbitrarily small 
values $J_\perp$ (for any $J$). 

\section{Discussion}

We have considered a model Hamiltonian motivated in part by the cuprate superconductors, and have
examined the formation of two-electron bound states in the dilute limit. Work on
related Hamiltonians \cite{lin,basu,pethukov} and on other model Hamiltonians motivated
by the cuprates \cite{marsiglio} also followed this approach in the hope of better understanding
the complicated many-electron states of these highly correlated systems.

This paper analyzed the formation of electron pair bound states in a 3$d$ lattice by comparing the 
minimum spin-exchange interaction that is necessary to allow for the formation of bound 
states for isotropic hopping and exchange, and for the complete suppression of interplanar hopping. 
We found that the minimum attractive interaction required is infinitesimal for the case where 
$t_\perp$ is zero, and hence the critical spin coupling for exchange-coupled planes is
\begin{equation}
	J_c = 0
\end{equation}

The question remains of trying to extrapolate with this very simple model
to the physics of two-electron bound state formation of the cuprate superconductors
(albeit in the dilute limit for each plane). A 3$d$ $t-J-U$ model that takes 
into account the correct anisotropies of the hopping and superexchange, and for which
$t_\perp$ is very small could indeed yield a two-electron bound state 
solution for very small values of $J_\perp$ {\em if} the limit of $J_c$ is continuous
as $t_\perp \rightarrow 0$. However, in general $J_{\perp,c}$ will depend on both
$t_\perp$ and $J_\parallel$, and the present study has not 
examined the parameter space associated with this much more difficult (numerically, that is) problem.
(Note from equation (\ref{eqnJz}) that there is no dependence on $J_\parallel$ when $t_\perp = 0$,
and for isotropic hopping and superexchange there is only dimensionless ratio, $J/t$.)

Our results could be of interest to the study of ultra-cold atoms - see reference \cite{coldatomsH} 
and references therein for the connection of such systems to the kind of model Hamiltonians studied
in connection with the cuprate superconductors. That is, in such systems one can tune the 
interactions, and with optical lattices the geometries, for such model Hamiltonians, so perhaps
the limit that was examined in this paper can be created in such system.

The above result necessarily leads to the question of the superconducting properties for such a model
Hamiltonian in the dilute limit. As shown in \cite{mraz}, and as found for the case of exchange-coupled 
chains~\cite{basu2}, weak (or in our case zero) interplanar electronic hopping leads to a large density 
of states at half filling (the density of states for a 2$d$ square lattice diverges logarithmically
at the middle of band), necessarily leading to large increases in $T_c$ when compared to the
situation found when the hopping frequency is isotropic. Therefore, using a weak-coupling BCS
approach one indeed expects for this model to show a large increase in $T_c$ relative to the 3$d$
lattice. What happens in the strong coupling limit remains a topic for conjecture.
In contrast to this result, note that 
for exchange-coupled chains~\cite{basu2} the electronic density of states is in fact divergent at 
the band edges, and thus not surprisingly one finds the greatest enhancement of $T_c$ at very low fillings.

\ack{We thank S. Basu for earlier discussions on this problem. This work was supported in part by the 
NSERC of Canada.}

\end{document}